\documentclass{webofc}
\usepackage[varg]{txfonts}   

\usepackage{bm}
\usepackage{comment}
\usepackage[subrefformat=parens]{subcaption}

\usepackage{xcolor}

\begin{document}
\title{
\vspace*{-23mm}\hspace{8.41cm} 
\small{\texttt{YITP-23-163, J-PARC-TH-0301}} 
\vspace*{10mm} \\
{\fontsize{14pt}{12pt}\selectfont 
Electromagnetic probes for critical fluctuations of phase transitions in dense QCD}
}
%

\author{
\firstname{Toru} \lastname{Nishimura}\inst{1,2}
\fnsep\thanks{\email{nishimura@kern.phys.sci.osaka-u.ac.jp}} 
\and
\firstname{Masakiyo} \lastname{Kitazawa}\inst{2,3}
\and
\firstname{Teiji} \lastname{Kunihiro}\inst{2}
}

\institute{
Department of Physics, Osaka University, 
560-0043, Toyonaka, Osaka, Japan
\and
Yukawa Institute for Theoretical Physics, Kyoto University, 
606-8502, Kyoto, Japan
\and
J-PARC Branch, KEK Theory Center, Institute of Particle and Nuclear Studies, 
KEK, 319-1106, Tokai, Ibaraki, Japan
}

\abstract{%
We study how the dilepton production rates and electric conductivity are affected 
by the phase transition to color superconductivity and the QCD critical point.
Effects of the soft modes associated with these phase transitions 
are incorporated through the photon self-energy 
called the Aslamazov-Larkin, Maki-Thompson, and density-of-states terms, 
which are responsible for the paraconductivity in metallic superconductors.
We show that anomalous enhancements of 
the production rate in the low energy/momentum region
and the conductivity occur around the respective critical points.
}
\maketitle

\section{Introduction}
\label{Introduction}
In the high baryon-density region of Quantum Chromodynamics (QCD),
rich phase structures are expected to exist.
They will be revealed
by various experimental programs of relativistic heavy-ion collisions (HIC) 
such as the beam-energy scan program at RHIC, HADES and NA61/SHINE, 
as well as the future experiments at FAIR, NICA and J-PARC-HI. 
In this proceeding, we discuss the possible observability of 
the color superconducting phase transition (CSC-PT) 
and the QCD critical point (QCD-CP) in these experiments 
through an anomalous enhancement of the dilepton production rate (DPR) 
caused by the soft modes associated with these 
phase transitions~\cite{Nishimura:2022mku,Nishimura:2023oqn}.
We also study the electric conductivity that is related to the low-energy limit of the DPR.
We calculate them by extending the theory of metallic superconductors~\cite{Larkin:book}.
It is shown that the soft modes lead to an anomalous enhancement 
of the DPR at low energy/momentum region 
and the conductivity near each phase transition.
We argue that these enhancements are used 
for experimental observables to verify the existence 
of the second-order phase transition on the QCD phase diagram~\cite{Nishimura:2023not}.

\section{Formalism}
\label{Formalism}
To study the dense medium around the 2-flavor superconductivity (2SC) and QCD-CP at nonzero temperature ($T$) and quark chemical potential ($\mu$),
we employ the 2-flavor NJL model
\begin{align}
  \mathcal{L} &= \bar{\psi} i ( \gamma^\mu \partial_\mu - m ) \psi 
  + \ G_S [(\bar{\psi} \psi)^2 + (\bar{\psi} i \gamma_5 \tau_i \psi)^2]
  + \ G_D (\bar{\psi} i \gamma_5 \tau_2 \lambda_A \psi^C)(\bar{\psi}^C i \gamma_5 \tau_2 \lambda_A \psi),
  \label{eq:lagrangian}
\end{align}
where $\tau_{i=1,2,3}$ is the Pauli matrix for the flavor $SU(2)_f$,
$\lambda_{A=2,5,7}$ is the antisymmetric components 
of the Gell-Mann matrices for the color $SU(3)_c$ and 
$\psi^C (x) \equiv i \gamma_2 \gamma_0 \bar{\psi}^T (x)$.
We use the scalar coupling $G_S=5.50~\rm{GeV^{-2}}$ and 
three-momentum cutoff $\Lambda=631\rm{MeV}$, 
which gives the pion mass $m_{\pi} = 138~\rm{MeV}$ 
and pion decay constant $f_{\pi} = 93~\rm{MeV}$
at the current quark mass $m = 5.5~{\rm MeV}$~\cite{Hatsuda:1994pi},
while the diquark coupling $G_D$ is treated as a free parameter.

It is known that the fluctuations of the diquark and chiral condensates
form collective modes with significant strength 
near the 2SC-PT~\cite{Kitazawa:2005vr} 
and the QCD-CP~\cite{Fujii:2004jt}, respectively.
The collective mode associated with the second-order 
phase transition is called the soft mode.
In the random-phase approximation, the propagators of the respective soft modes 
in the imaginary-time formalism are evaluated as
\begin{align}
\tilde{\Xi}_D(k) &= \frac{1}{{G_D}^{-1}+\mathcal{Q}_D(k)},\quad
\mathcal{Q}_D (k) = 
 \begin{minipage}[h]{0.1\linewidth}
 \centering
 \includegraphics[keepaspectratio, scale=0.034]{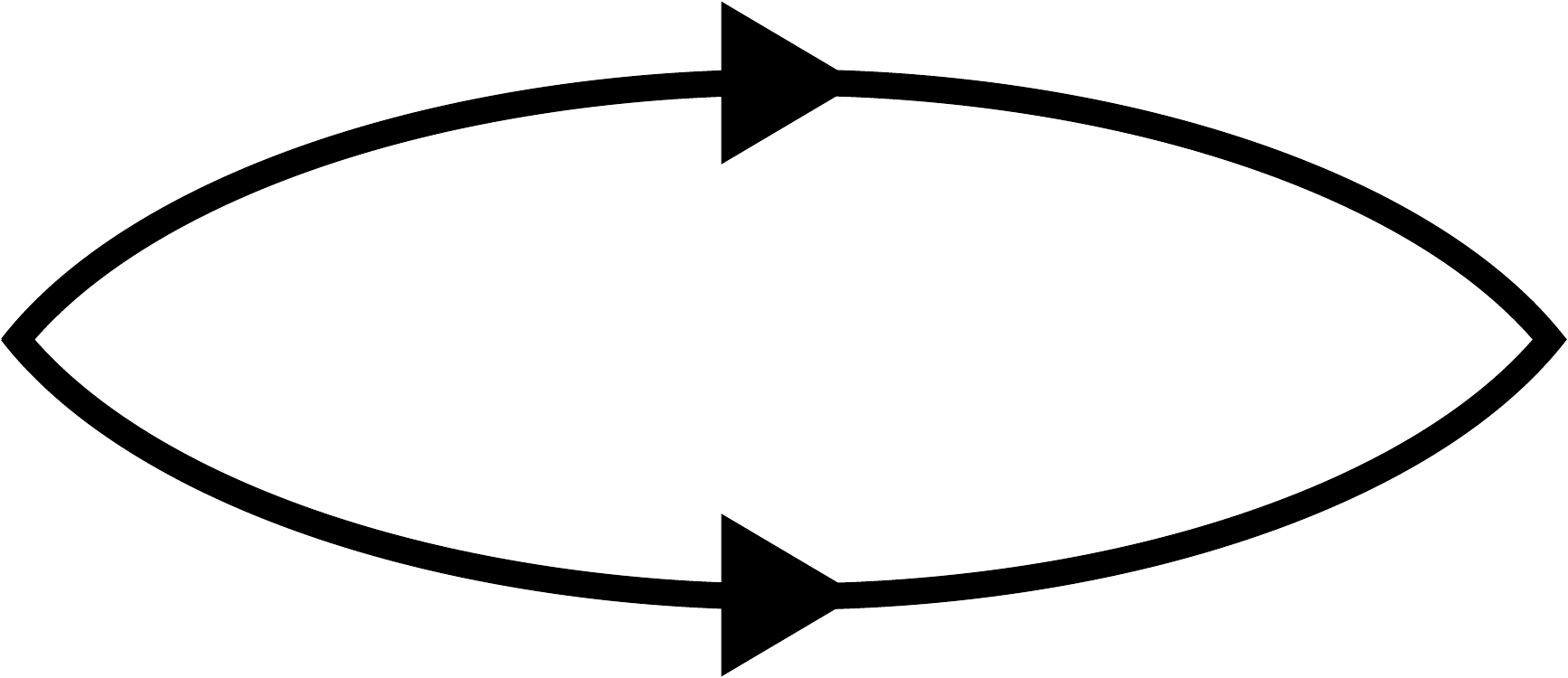}
 \end{minipage}
=-2 N_f (N_c-1) \int_p {\rm Tr} [\mathcal{G}_0 (k-p) \mathcal{G}_0(p)],
\label{eq:softmode_CSC} 
\\
\tilde{\Xi}_S(k) &= \frac{1}{{G_S}^{-1}+\mathcal{Q}_S(k)},\quad
\mathcal{Q}_S (k) = 
 \begin{minipage}[h]{0.1\linewidth}
 \centering
 \includegraphics[keepaspectratio, scale=0.034]{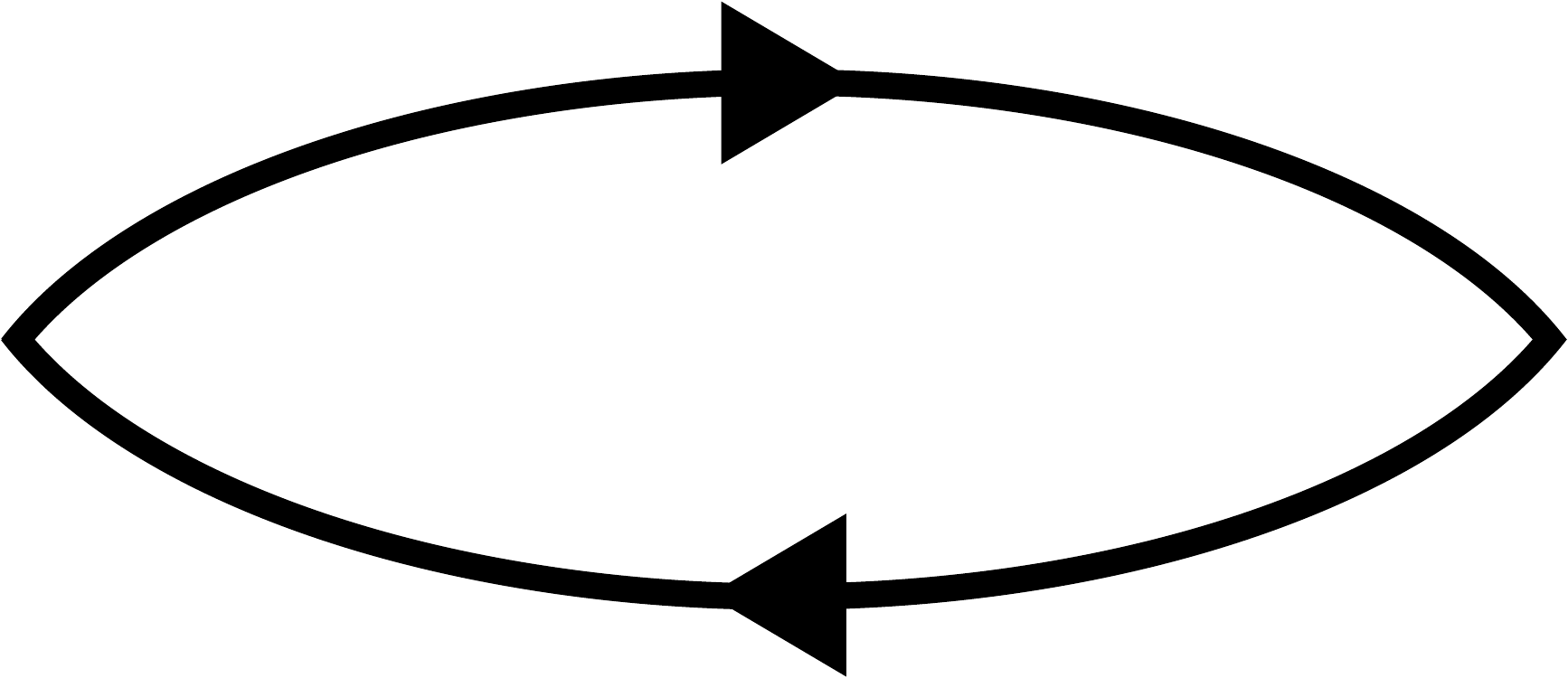}
 \end{minipage}
=2 N_f N_c \int_p {\rm Tr} [\mathcal{G}_0 (p-k) \mathcal{G}_0(p)],
\label{eq:softmode_QCDCP} 
\end{align}
where $\mathcal{Q}_D(k)=\mathcal{Q}_D(\bm{k}, i\nu_n)$ 
and $\mathcal{Q}_S(k)=\mathcal{Q}_S(\bm{k}, i\nu_n)$ are
the one-loop $qq$ and $q\bar{q}$ correlation function, 
respectively, with the free quark propagator
$\mathcal{G}_0(p)=\mathcal{G}_0(\bm{p}, i\omega_m) = 
1/[(i\omega_m + \mu)\gamma_0 - \bm{p} \cdot \bm{\gamma} - M]$, 
the Matsubara frequency for fermions (bosons) $\omega_m$ ($\nu_n$), the trace over Dirac indices ${\rm Tr}$, 
and $\smallint_p = T\sum_m \int d^3p/(2\pi)^3$.
The retarded propagator $\Xi^R_\gamma (\bm{k}, \omega)$
($\gamma = D, S$) is obtained by the analytic continuation 
$i\nu_n \rightarrow \omega+i\eta$.

We remark that the propagators~\eqref{eq:softmode_CSC} and~\eqref{eq:softmode_QCDCP} satisfy the Thouless criterion, i.e.
${\Xi^{R}_\gamma}^{-1} (\bm{0}, 0) = 0$ at the respective critical points.
The criterion shows that a pole of the propagator 
approaches the origin of the complex energy plane 
as $T$ and $\mu$ goes toward the critical points,
and hence the fluctuations of the condensates 
become soft near the critical points.

The DPR is related to the retarded photon self-energy $\Pi^{R \mu\nu}(k)$ as 
\begin{align}
\frac{d^4\Gamma}{d^4k} = - \frac{\alpha}{12\pi^4} 
\frac{1}{k^2} \frac{1}{e^{\omega/T}-1} g_{\mu\nu}  
{\rm Im} \Pi^{R \mu\nu} (k) ,
\label{eq:DPR}
\end{align}
where $\alpha$ is the fine structure constant. 
The elecrtric conductivity is given in terms of the spectral density 
$\rho(\omega) = -\sum_i {\rm Im} \Pi^{R ii} (\bm{0}, \omega)$ as
\begin{align}
\sigma = \frac{1}{3} \frac{\partial \rho(\omega)}{\partial \omega} \bigg|_{\omega=0}~.
\label{eq:sigma}
\end{align}


To construct the photon self-energy
$\tilde{\Pi}^{\mu\nu}(k)$ that includes the soft modes,
we start from the lowest contribution of the soft modes 
to the thermodynamic potential 
$\Omega_\gamma = \smallint_p \ln [G_\gamma \tilde\Xi_\gamma^{-1} (p)]$,
which is the one-loop diagram of $\tilde{\Xi}_\gamma(p)$.
The self-energy is then obtained by attaching electromagnetic 
vertices to two points of quark lines in $\Omega_\gamma$.
From this procedure, one finds three types of diagrams
which are called the Aslamazov-Larkin (AL), 
Maki-Thompson (MT) and density-of-states (DOS) terms
in the theory of metallic superconductivity.
Each contribution to the photon self-energy is given by
\begin{align}
\tilde{\Pi}_{{\rm AL},D}^{\mu\nu} (k) &= 3 \int_q 
\tilde{\Gamma}_D^\mu(q, q+k) \tilde{\Xi}_D(q+k) 
\tilde{\Gamma}_D^\nu(q+k, q) \tilde{\Xi}_D(q),
\label{eq:AL_D} 
\\
\tilde{\Pi}_{{\rm MT (DOS)},D}^{\mu\nu} (k) &= 3 \int_q 
\tilde{\Xi}_D(q) \mathcal{R}_{{\rm MT (DOS)},D}^{\mu\nu}(q, k), 
\label{eq:MTDOS_D} 
\\
\tilde{\Pi}_{{\rm AL},S}^{\mu\nu} (k) &= \sum_{f=u,d} \int_q 
\tilde{\Gamma}_f^\mu(q, q+k) \tilde{\Xi}_S(q+k) 
\tilde{\Gamma}_f^\nu(q+k, q) \tilde{\Xi}_S(q),
\label{eq:AL_S} 
\\
\tilde{\Pi}_{{\rm MT (DOS)},S}^{\mu\nu} (k) &= \sum_{f=u,d} \int_q 
\tilde{\Xi}_S(q) \mathcal{R}_{{\rm MT (DOS)},f}^{\mu\nu}(q, k), 
\label{eq:MTDOS_S} 
\end{align}
where the vertices $\tilde{\Gamma}_D^\mu(q, q+k)$, 
$\mathcal{R}^{\mu\nu}_D(q, k)
=\mathcal{R}_{{\rm MT},D}^{\mu\nu}(q, k)
+\mathcal{R}_{{\rm DOS},D}^{\mu\nu}(q, k)$,
$\tilde{\Gamma}_f^\mu(q, q+k)$, and
$\mathcal{R}^{\mu\nu}_f(q, k)
=\mathcal{R}_{{\rm MT},f}^{\mu\nu}(q, k)
+\mathcal{R}_{{\rm DOS},f}^{\mu\nu}(q, k)$ 
satisfy the Ward-Takahashi (WT) identities 
\begin{align}
k_\mu \tilde{\Gamma}_D^\mu (q, q+k) 
&= (e_u+e_d) [\mathcal{Q}_D(q+k)-\mathcal{Q}_D(q)], 
\label{eq:AL-vertex_D} 
\\
k_\mu \mathcal{R}_D^{\mu\nu} (q, k) 
&= (e_u+e_d) [\tilde{\Gamma}_D^\nu (q-k, q)-\tilde{\Gamma}_D^\nu (q, q+k)], 
\label{eq:MTDOS-vertex_D}
\\
k_\mu \tilde{\Gamma}_f^\mu (q, q+k) 
&= -e_f [\mathcal{Q}_S(q+k)-\mathcal{Q}_S(q)], 
\label{eq:AL-vertex_S} 
\\
k_\mu \mathcal{R}_f^{\mu\nu} (q, k) 
&= -e_f [\tilde{\Gamma}_f^\nu (q-k, q)-\tilde{\Gamma}_f^\nu (q, q+k)], 
\label{eq:MTDOS-vertex_S}
\end{align}
with $e_u=2|e|/3$ ($e_d=-|e|/3$) being the electric charge of the up (down) quark.
The total photon self-energy is then given by $\tilde{\Pi}^{\mu\nu} (k) = \tilde{\Pi}_{\rm free}^{\mu\nu} (k) + \tilde{\Pi}_D^{\mu\nu} (k) +\tilde{\Pi}_S^{\mu\nu} (k) $ with
\begin{align}
\tilde{\Pi}_D^{\mu\nu} (k) 
= 
& \tilde{\Pi}_{{\rm AL},D}^{\mu\nu} (k)
+ \tilde{\Pi}_{{\rm MT},D}^{\mu\nu} (k)
+ \tilde{\Pi}_{{\rm DOS},D}^{\mu\nu} (k) ,
\\
\tilde{\Pi}_S^{\mu\nu} (k) 
= 
& \tilde{\Pi}_{{\rm AL},S}^{\mu\nu} (k)
+ \tilde{\Pi}_{{\rm MT},S}^{\mu\nu} (k)
+ \tilde{\Pi}_{{\rm DOS},S}^{\mu\nu} (k),
\label{eq:Pi} 
\end{align}
where $\tilde{\Pi}_{\rm free}^{\mu\nu} (k)$ is the contribution of the massless free quark gases.
One can explicitly check that $\tilde{\Pi}^{\mu\nu} (k)$ satisfies the WT identity $k_\mu\tilde{\Pi}^{\mu\nu} (k)=0$
using Eqs.~\eqref{eq:AL-vertex_D}--\eqref{eq:MTDOS-vertex_S}.

\section{Numerical results and summary}
\label{Formalism}

\begin{figure}[tbp]
	\begin{tabular}{cc}
      \begin{minipage}[t]{0.47\hsize}
        \centering
        \includegraphics[keepaspectratio, scale=0.45]{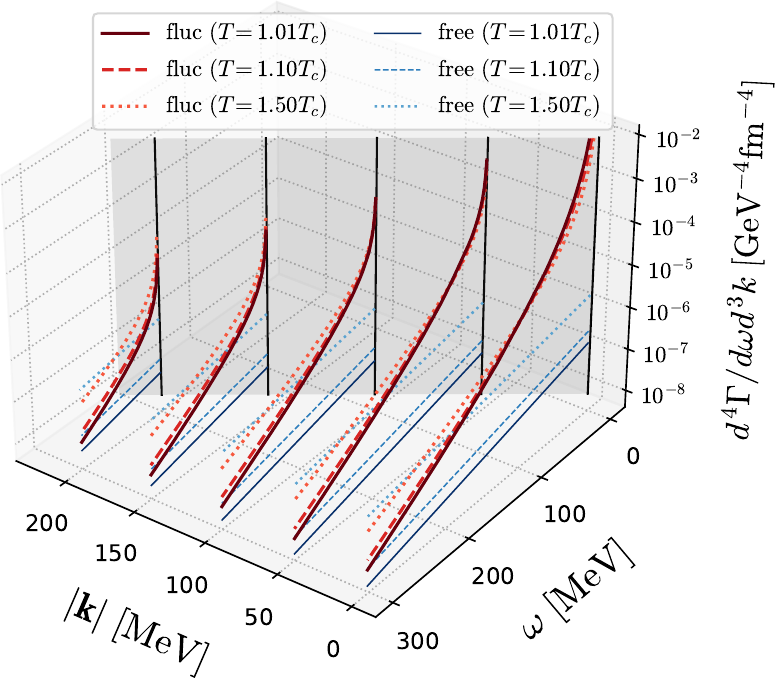}
      \end{minipage} &
      \begin{minipage}[t]{0.47\hsize}
        \centering
        \includegraphics[keepaspectratio, scale=0.45]{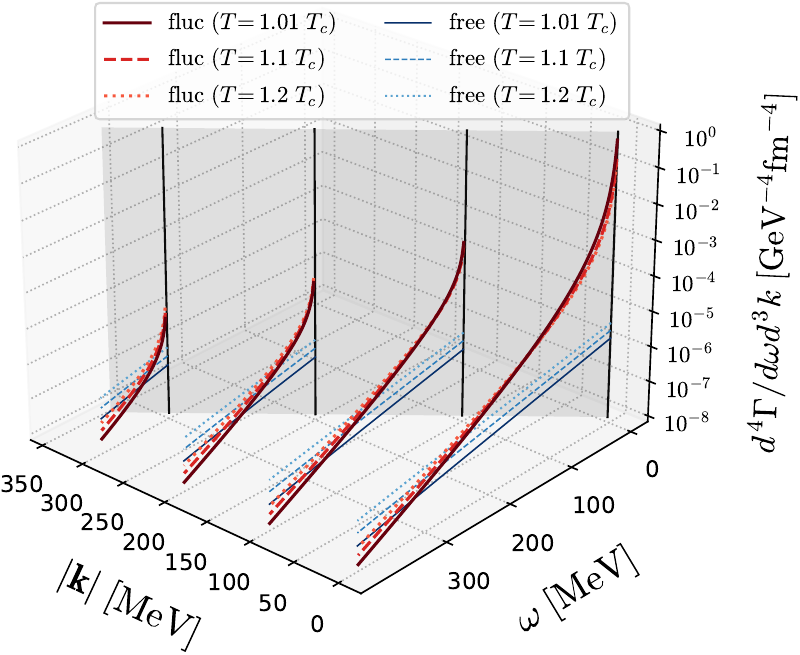}
      \end{minipage}
	\end{tabular}
\caption{
Dilepton production rates per unit energy $\omega$ and momentum $\bm{k}$
above $T_c$ of the 2SC at $\mu = 350~{\rm MeV}$ 
(left)~\cite{Nishimura:2022mku}
and of the QCD-CP at $\mu = \mu_{\rm CP}$ 
(right)~\cite{Nishimura:2023oqn} with $G_D = 0.7G_S$. 
The thick (thin) lines are the contribution of the soft modes (the massless free quark gases). 
}
\vspace{-5mm}
\label{fig:RATE}
\end{figure}

The left panel of Fig.~\ref{fig:RATE} shows the DPR 
per unit energy $\omega$ and momentum $\bm{k}$
near the 2SC-PT for $T/T_c=1.01$, $1.1$ and $1.5$ at $\mu=350~{\rm MeV}$ 
and $G_D = 0.7G_S$ with
the critical temperature $T_c =42.94~{\rm MeV}$~\cite{Nishimura:2022mku}.
The thick lines are the contributions from the soft modes, 
while the thin lines are the ones of the free quark gas.
One sees that the DPR from the soft modes is anomalously enhanced 
at small $\omega$ and $\bm{k}$ region in comparison 
with the free quark gas for $T\lesssim1.5T_c$,
and this enhancement is more pronounced as $T$ approaches $T_c$.
This result is expected from the properties of the soft modes.
Shown in the right panel of Fig.~\ref{fig:RATE} is the DPR near the QCD-CP for 
$T/T_{\rm CP}=1.01$, $1.1$ and $1.2$ 
at $\mu=\mu_{\rm CP}$
with the location of the QCD-CP $(T_{\rm CP}, \mu_{\rm CP}) = (46.712, 329.34)~{\rm MeV}$~\cite{Nishimura:2023oqn}.
One finds that the anomalous enhancement of 
the DPR is observed similarly to the case of the 2SC.

\begin{figure}[t]
\centering
\includegraphics[keepaspectratio, scale=0.56]{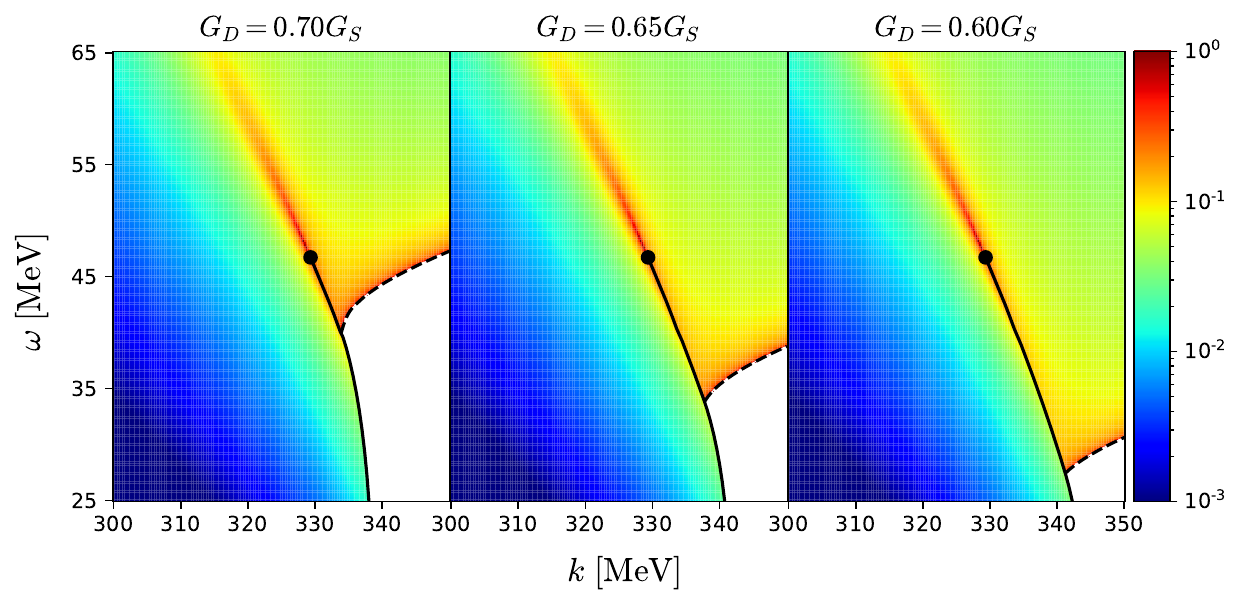}
\caption
{
Contour maps of the electric conductivity $\sigma/T$ on the $T$--$\mu$ plane 
around the QCD-CP and 2SC-PT with $G_D/G_S= 0.70$, $0.65$ and $0.60$.
The solid (dashed) line is the first-order (second-order) phase transition.
}
\vspace{-5mm}
\label{fig:sigma}
\end{figure}

Figure~\ref{fig:sigma} shows the contour maps 
of the electric conductivity $\sigma/T$ on the $T$--$\mu$ plane 
for $G_D/G_S = 0.70$, $0.65$ and $0.60$
obtained from $\tilde\Pi^{\mu\nu}(k)$ with the effects of the 2SC-PT and QCD-CP.
One sees that the conductivity is enhanced around these phase transitions.
Since the effect of finite diquark condensate is not considered, 
our formalism is not applicable to the 2SC phase, i.e. the white region in the figure.

In Fig.~\ref{fig:sigma}, it is notable that the electric conductivity 
is enhanced at two distinct regions on the $T$--$\mu$ plane near the QCD-CP and 2SC-PT, 
whose separation is controlled by the value of $G_D$. 
Since the electric conductivity is related to the DPR at low-mass region 
through Eqs.~(\ref{eq:DPR}) and~(\ref{eq:sigma}), this result indicates 
that there are also two hot spots of the dilepton production on the QCD phase diagram.
In the beam-energy scan of the HIC where different regions on the $T$--$\mu$ plane 
can be investigated by varying the collision energy $\sqrt{s_{_{NN}}}$, 
this behavior would result in two distinct enhancements 
of the dilepton yield as a function of $\sqrt{s_{_{NN}}}$, 
as recently demonstrated in Ref.~\cite{Nishimura:2023not}.
It is quite interesting to pursue such behavior in the HIC, as it becomes 
strong collateral evidence for the existence of both the QCD-CP and 2SC-PT.

In this proceeding, we studied how the soft modes of the 2SC-PT and QCD-CP
affect the DPR and electric conductivity around the phase transitions.
The modifications of the photon self-energy due to the soft modes are incorporated 
through the Aslamazov-Larkin, Maki-Thompson, and density-of-states terms,
in accordance with the WT identity of the photon self-energy.
The characteristic structure of the enhancement of the DPR 
on the QCD phase diagram found in this study would allow us 
to detect the 2SC-PT and QCD-CP in future HIC experiments.

%
%

\end{document}